\documentclass[aps,prb,twocolumn,superscriptaddress]{revtex4}

\usepackage{graphicx}% Include figure files

\begin{document}

%\preprint{}

\title{ Coupling-dependent rates of energy transfers from photoexcited Mott insulators to lattice vibrations }

% Please use the appropriate macro for each type of information

\author{ Kenji Yonemitsu }
\email[]{kxy@ims.ac.jp}
%\thanks{}
\affiliation{ Institute for Molecular Science, Okazaki 444-8585, Japan }
\affiliation{ Department of Functional Molecular Science, Graduate University for Advanced Studies, Okazaki 444-8585, Japan }
\author{ Nobuya Maeshima }
%\email[]{}
%\thanks{}
\affiliation{ Institute of Materials Science, University of Tsukuba, Tsukuba 305-8573, Japan }

\date{\today}

\begin{abstract}
Photoexcited states are relaxed by transferring energy to the environments. In order to study which coupling allows fast energy transfer to lattice vibrations in correlated electron systems, we calculate the time evolutions of the kinetic energies of different types and frequencies of lattice vibrations. The one-dimensional half-filled Hubbard model is augmented with electron-lattice couplings that modulate transfer integrals, site energies, and Coulomb repulsion strengths. The time-dependent Schr\"odinger equation is solved for exact many-electron wave functions, and the classical equation of motion for the lattice displacements. In order to transfer energy to classical lattice vibrations that modulate transfer integrals or site energies, the translational invariance must be broken to give optical activity to an electronic excitation with wave number $ \pi $ and to these lattice vibrations. On the other hand, a certain amount of energy is always transferred to lattice vibrations that modulate Coulomb repulsion strengths, irrespective of the symmetry of the ground state, as long as the corresponding electron-lattice couplings are present. In strongly correlated electron systems, these couplings can be strong, although they are usually insignificant because their effects on the equilibrium properties can be absorbed into redefinition of Coulomb repulsion strengths. We will discuss competition or collaboration between energy transfer pathways through different types of electron-lattice couplings. 
\end{abstract}

\pacs{78.20.Bh, 71.10.Fd, 71.38.-k, 78.47.J-}
% 78.20.Bh Theory, models, and numerical simulation  
% 71.10.Fd Lattice fermion models (Hubbard model, etc.)  
% 71.38.-k Polarons and electron-phonon interactions (see also 63.20.Kr)
% 78.47.J- Ultrafast pump/probe spectroscopy (< 1 psec) 
% 71.30.+h Metal-insulator transitions and other electronic transitions  
% 63.20.kd Phonon-electron interactions  
% 71.45.Lr Charge-density-wave systems (see also 75.30.Fv Spin-density waves) 
\keywords{photoinduced phase transition, relaxation, electron correlation}

\maketitle

\section{Introduction}

Photoinduced phase transitions have attracted much attention as intriguing cooperative phenomena in nonequilibrium conditions. \cite{yonemitsu_jpsj06,yonemitsu_pr08} The electronic properties caused by many electrons are macroscopically changed. Especially in strongly correlated electron systems, they are changed often on an ultrafast time scale. The density of photons needed for such a change is usually much lower than that of electrons involved because the energy supplied by photoirradiation is much higher than thermal energies. 

Quasi-one-dimensional Mott-Hubbard insulators such as halogen-bridged nickel-chain compounds \cite{kishida_n00} and Sr$_2$CuO$_3$ \cite{kishida_n00,ogasawara_prl00} have been expected as nonlinear optical materials with large nonlinear coefficients and fast response times. Ultrafast recovery times are desirable for optical high-speed switching devices. Then, fast energy dissipation from the photoexcited state into the environments is required. For Sr$_2$CuO$_3$, the observed photoinduced absorption is to an even-parity two-photon state that occurs immediately above the absorption edge, which is theoretically explained on the basis of the one-dimensional extended Hubbard model with alternating site energies. \cite{ogasawara_prl00} In such one-dimensional correlated electron systems, the spin-charge separation plays a crucial role in enhancing the nonlinear optical response. \cite{mizuno_prb00}

Halogen-bridged nickel-chain compounds are known to exhibit a photoinduced insulator-to-metal transition and ultrafast decay of the photoinduced state. \cite{iwai_prl03} Their photoinduced metallic properties are theoretically studied in the one-dimensional Hubbard model. \cite{maeshima_jpsj05} The difference between the photoinduced state in the nickel-chain compounds and that in the copper-oxide chains is experimentally clarified by comparisons between their optical conductivity spectra and photoconductivity spectra. \cite{ono_prb04}

Quasi-one-dimensional Mott-Hubbard insulators are often accompanied by the dimerization that alternates transfer integrals to produce a spin-Peierls phase. Photoirradiation of such insulators as K-tetracyanoquinodimethane (K-TCNQ) are known to exhibit melting of the dimerization, which is called a photoinduced inverse spin-Peierls transition. \cite{koshihara_prb91} Ultrafast photoinduced melting of the spin-Peierls phase is observed with a very short decay time of photocarriers. \cite{okamoto_prl06} In  bis(ethylenedithio)tetrathiafulvalene-difluorotetracyanoquinodimethane (ET-F$_2$TCNQ), which is regular without dimerization, photocarrier doping is concentrated on a Drude component and a midgap state is never formed owing to the spin-charge separation. \cite{okamoto_prl07} For a larger photoexcitation, the decay time of the metallic state is significantly shorter, suggesting that electron-electron scattering plays an important role. 

Recently, ultrafast charge dynamics in K-TCNQ, Rb-TCNQ, and ET-F$_2$TCNQ with different magnitudes of electron-lattice interactions are compared. \cite{uemura_jpsj08} In K-TCNQ and Rb-TCNQ, photocarriers are localized as polarons with about 70 fs and recombine with a few ps. In ET-F$_2$TCNQ, photocarriers lead to a metallic state and decay with about 200 fs. These differences in the charge dynamics are reasonably interpreted as due to different magnitudes of electron-lattice interactions. The photoinduced inverse spin-Peierls transition is also theoretically studied in the one-dimensional dimerized Hubbard model \cite{maeshima_prb06} and in its extension to include nearest-neighbor repulsion. \cite{maeshima_jpsj07} To understand photoinduced midgap states, however, electron-lattice interactions and the rapid formation of polaronic states are required for K-TCNQ. \cite{maeshima_jpsj08} 

In correlated electron systems on the one-dimensional regular lattice at half filling, the spin-charge separation holds in the low-energy limit and almost holds even in photoexcited states. Consequently, electron-lattice interactions play an important role in the decay of photocarriers. In studying the relaxation of photoexcited states in correlated electron systems, it is numerically hard to observe the time evolution of the conductivity spectrum or the Drude weight, but it is easy to observe the time-dependent kinetic energies of phonons. In this paper, we then focus on the energy transfers from photoexcited electron systems to phonons. 

Quite recently, enhancement of phonon effects on an exciton has been reported in photoexcited states of one-dimensional Mott insulators on the basis of a density matrix renormalization group calculation for the one-dimensional extended Hubbard-Holstein model. \cite{matsueda_prb08} Here, quantum phonons are incorporated and play an important role because the holon and the doublon are dressed with quantum phonons. For photoinduced dynamics near a quantum critical point, where quantum phonon fluctuations are essential, coherence is shown to be enhanced within the quantum Blume-Emery-Griffiths model. \cite{yonemitsu_prb08a} In this paper, however, we will limit ourselves to classical phonons, which can suggest that electron-lattice couplings that are usually ignored can lead to ultrafast decay of the photoinduced state in correlated electron systems. 

We here briefly summarize the relation between quantum and classical phonons. Phonons originally behave quantum-mechanically. When the average number of phonons at each site is much larger than the fluctuating component, they behave classically. If phonon energies are not negligibly small compared with electronic kinetic energies, phonons follow the motion of electrons to produce slowly moving polarons. The quantum nature of phonons must be retained to produce polarons in a translation invariant manner. With decreasing phonon energies, the motion of phonons is more retarded than that of electrons, and finally electrons feel as if phonons produce a quasi-static field. Phonons barely follow the motion of electrons to create hardly moving polarons. In the classical picture, polarons break the translational symmetry, but quantum fluctuations restore the symmetry by the linear combination of differently placed polarons. In reality, thermal fluctuations would help the localization of polarons. 

In K-TCNQ, photocarriers are regarded as converted into polarons \cite{okamoto_prl06,uemura_jpsj08} presumably by relaxing the displacements that modulate transfer integrals. \cite{maeshima_jpsj08} It is interpreted that photocarriers are localized and their recombination (i.e., the decay of photocarriers) is slowed down by the corresponding electron-lattice interaction. \cite{uemura_jpsj08} The fact that the materials with stronger electron-lattice interactions have the longer decay times suggests either that electron-electron interactions are mainly responsible for the decay process \cite{uemura_jpsj08} or that a new type of electron-lattice interactions compete with those responsible for the formation of polarons in determining the decay time. 

In this paper, we will suggest the latter possibility by introducing electron-lattice interactions that modulate Coulomb repulsion strengths. These interactions allow energy transfers to phonons, irrespective of the symmetry of the ground state. As electron-lattice interactions that modulate transfer integrals increase, the dimerization is enhanced, and the energy transfers to phonons that modulate Coulomb repulsion strengths are shown to decrease in this paper using classical phonons. This fact will survive quantum fluctuations of phonons. Numerical calculations for time evolution with quantum phonons on a reasonably large lattice at half filling are unfeasible and beyond the scope of this paper. 

\section{Model and Method}

In order to study the energy dissipation from a correlated electron system into the environments, we take a one-dimensional half-filled Peierls-Holstein-Hubbard model with another type of electron-lattice couplings that modulate on-site repulsion strengths, 
\begin{eqnarray}
H & = & 
-\sum_{j,\sigma} \left[ 
t_0-\sum_m \sqrt{ s_\alpha^{(m)} } \left( u^{(m)}_{j+1} - u^{(m)}_{j}  \right)
\right] \left( c^\dagger_{j,\sigma} c_{j+1,\sigma} \right.
\nonumber \\ & &  \left.
+ c^\dagger_{j+1,\sigma} c_{j,\sigma} 
\right ) -\sum_{j,m} 
\sqrt{ s_\beta^{(m)} } v^{(m)}_{j} (n_j-1) \nonumber \\ & &
+ \sum_j \left(
U -\sum_m \sqrt{ s_\gamma^{(m)} } w^{(m)}_{j}
\right) n_{j,\uparrow} n_{j,\downarrow}
 \nonumber \\ & &
+ \frac12 \sum_{j,m} \left( u^{(m)}_{j+1} - u^{(m)}_{j} \right)^2
+ \frac12 \sum_{j,m} v^{(m)2}_{j}
+ \frac12 \sum_{j,m} w^{(m)2}_{j}
\nonumber \\ & &
+ 2 \sum_{j,m} \frac{ \dot{u}^{(m)2}_{j} }{\omega_\alpha^{(m)2}} 
+ \frac12 \sum_{j,m} \frac{ \dot{v}^{(m)2}_{j} }{\omega_\beta^{(m)2}} 
+ \frac12 \sum_{j,m} \frac{ \dot{w}^{(m)2}_{j} }{\omega_\gamma^{(m)2}}  
\;, \label{eq:model}
\end{eqnarray}
where $ c^\dagger_{j,\sigma} $ ($c_{j,\sigma} $) creates (annihilates) an electron with spin $ \sigma $ at site $ j $, $ n_{j,\sigma} = c^\dagger_{j,\sigma} c_{j,\sigma} $, $ n_j = \sum_\sigma n_{j,\sigma} $, $ t_0 $ denotes the bare transfer integral, and $ U $ the bare on-site repulsion strength. As for electron-lattice couplings, we consider different types and frequencies of lattice vibrations. The lattice displacements $ u^{(m)}_{j} $, $ v^{(m)}_{j} $, and $ w^{(m)}_{j} $ modulate the transfer integral, the site energy, and the on-site repulsion strength, with coupling strengths, $ s_\alpha^{(m)} $, $ s_\beta^{(m)} $, and $ s_\gamma^{(m)} $, respectively, at bond or site $ j $. Here, $ s_\nu^{(m)} $ ($ \nu $=$ \alpha $, $ \beta $, $ \gamma $) are not smaller than zero, and the suffix $ m $ denotes modes having the bare phonon frequencies, $ \omega_\alpha^{(m)} $, $ \omega_\beta^{(m)} $, and $ \omega_\gamma^{(m)} $. The quantities $ \dot{u}^{(m)}_{j} $, $ \dot{v}^{(m)}_{j} $, and $ \dot{w}^{(m)}_{j} $ are the time derivatives of $ u^{(m)}_{j} $, $ v^{(m)}_{j} $, $ w^{(m)}_{j} $, respectively. For simplicity, these displacements are regarded as independent. 

We define each type of electron-phonon coupling strength by 
$ \lambda_\alpha \equiv \sum_m s_\alpha^{(m)} $, 
$ \lambda_\beta  \equiv \sum_m  s_\beta^{(m)} $, and 
$ \lambda_\gamma \equiv \sum_m s_\gamma^{(m)} $. 
As long as the lattice vibrations are treated classically, the ground state is given by $ \dot{u}^{(m)}_{j} $=$ \dot{v}^{(m)}_{j} $=$ \dot{w}^{(m)}_{j} $=0. In this case, the ground state is determined not by the distribution of $ \{ s_\alpha^{(m)} \} $, $ \{ s_\beta^{(m)} \} $, or $ \{ s_\gamma^{(m)} \} $, but by their sums, $ \lambda_\alpha $, $ \lambda_\beta $, and $ \lambda_\gamma $, as derived in a straightforward manner from the Hellmann-Feynman theorem. Therefore, we will specify the electron-phonon coupling strengths by $ \lambda_\alpha $, $ \lambda_\beta $, and $ \lambda_\gamma $ in the following, although the dynamics induced by any excitation depends on the distribution of couplings. 

In this paper, we consider many phonon modes of different frequencies for each type of electron-phonon couplings to reduce coherent oscillations of energy exchange between the electronic and lattice systems. Therefore, we take a wide distribution of phonon frequencies: $ \omega_\nu^{(m)} = \omega_\nu \times m/100 $ for $ \nu $=$ \alpha $, $ \beta $, $ \gamma $ and $ m $=1, $ \cdots $, 100 with $ \omega_\nu $ being the maximum phonon energy $ \omega_\nu \equiv \max_m \omega_\nu^{(m)} $. Furthermore, we take a uniform distribution of couplings: $ s_\nu^{(m)} = s_\nu $. The number 100 is more than realistic values even for complex molecular materials, but it is so taken as to be large enough in that the numerical results are almost unchanged if we take 10$^4$ instead. Although phonons are treated classically in this paper, we employ exact many-electron wave functions for the ground state and the photoinduced dynamics. 

Photoexcitations are introduced by adding 
\begin{eqnarray} &  & 
-\sum_{j,\sigma} \left[ 
t_0-\sum_m \sqrt{ s_\alpha^{(m)} } \left( u^{(m)}_{j+1} - u^{(m)}_{j}  \right)
\right] \nonumber \\ & & \times
\left\{ \exp\left[ -\frac{iea}{\hbar} \int dt E(t) \right]-1 \right\}
c^\dagger_{j,\sigma} c_{j+1,\sigma} + \mathrm{H.c.}
\;, \label{eq:field}
\end{eqnarray}
to Eq.~( \ref{eq:model}), where $ e $ is the absolute value of the electronic charge, $ a $ is the lattice spacing, and $ e $, $ a $, and $ \hbar $ are set to be unity. The time-dependent electric field $ E(t) $ is given by $ E(t) = E_\mathrm{ext} \sin \omega_\mathrm{ext} t $, with amplitude $ E_\mathrm{ext} $ and frequency $  \omega_\mathrm{ext} $ for $ 0 < t < T_\mathrm{irr} $ [$ E(t) $ is zero otherwise.] with $ T_\mathrm{irr} = 2\pi N_\mathrm{ext}/\omega_\mathrm{ext} $ and integer $ N_\mathrm{ext} $. 

The time-dependent Schr\"odinger equation for the exact many-electron wave function $ \mid \! \psi (t) \rangle $ is numerically solved by 
\begin{equation}
\mid \! \psi (t+dt) \rangle \simeq 
\exp \left[-\frac{ i }{ \hbar } dt H \left(t+\frac{ dt }{ 2 }\right) \right]
\mid \! \psi (t) \rangle 
\;,
\end{equation}
where the time evolution operator is expanded as 
\begin{equation}
\exp \left[-\frac{ i }{ \hbar } dt H \left(t+\frac{ dt }{ 2 }\right) \right] = 
\sum_{n=0}^{\infty} \frac{1}{n!} 
\left[-\frac{ i }{ \hbar } dt H \left(t+\frac{ dt }{ 2 }\right) \right]^n 
\;,
\end{equation}
with time slice $ dt $=10$^{-3}$ to the ($ n $=)15th order and by checking the conservation of the norm and of the total energy for $ t > T_{\text{irr}} $. This method is the same as in Refs.~\onlinecite{yonemitsu_prb07,onda_prl08}, but we here use a much smaller value for $ dt $ to raise the precision. The classical equation of motion for the lattice displacements is solved by the leapfrog method, where the force is derived from the Hellmann-Feynman theorem. 

To observe the energy dissipation into the environments, we calculate the time evolution of the kinetic energy of each type of phonons, 
\begin{equation}
E_{\rm kin}(\{u\})=\left\langle 
2 \sum_{j,m} \frac{ \dot{u}^{(m)2}_{j} }{\omega_\alpha^{(m)2}} 
\right\rangle 
\;, \label{eq:kinetic_u}
\end{equation}
\begin{equation}
E_{\rm kin}(\{v\})=\left\langle 
\frac12 \sum_{j,m} \frac{  \dot{v}^{(m)2}_{j} }{\omega_\beta^{(m)2}} 
\right\rangle 
\;, \label{eq:kinetic_v}
\end{equation}
\begin{equation}
E_{\rm kin}(\{w\})=\left\langle 
\frac12 \sum_{j,m} \frac{ \dot{w}^{(m)2}_{j} }{\omega_\gamma^{(m)2}} 
\right\rangle 
\;, \label{eq:kinetic_w}
\end{equation}
and compare the time evolution of the total energy, $ E_\mathrm{tot} = \langle H \rangle $. In the following, the distortion amplitudes in the initial ground state are defined by the dimensionless quantities, 
\begin{equation}
u_\mathrm{GS} \equiv \mid u^{(m)}_{j} \mid / \sqrt{ s_\alpha^{(m)} }
\;, \label{eq:dist_amp_u}
\end{equation}
\begin{equation}
v_\mathrm{GS} \equiv \mid v^{(m)}_{j} \mid / \sqrt{ s_\beta^{(m)} }
\;, \label{eq:dist_amp_v}
\end{equation}
\begin{equation}
w_\mathrm{GS} \equiv \mid w^{(m)}_{j} \mid / \sqrt{ s_\gamma^{(m)} }
\;. \label{eq:dist_amp_w}
\end{equation}

\section{Results}

In this paper, we use $ t_0 $=1 and $ U $=6 for Mott insulators and $ \omega_\mathrm{ext} $=3, which is slightly above the lowest optical excitation energy. We take $ N_\mathrm{ext} $=20, so that $ T_\mathrm{irr} $=41.9, which corresponds to 27.5 fs (275 fs) if $ t_0 $ is 1 eV (0.1 eV). Furthermore, we use $ E_\mathrm{ext} $=1, which is not so large in that the increment in the total energy ranges about 1.5 to 3 (i.e., 0.5 to 1 photon) in the present 12-site system. Although it is not easy to obtain general results on the differences between Mott and band insulators, we first compared the energy transfer to phonons in a Mott insulator and that in a band insulator, where the bare transfer integral $ t_0 $ in Eq.~( \ref{eq:model}) is replaced by $ t_0 -(-1)^j \delta t $, $ \delta t $ is set at 0.75, and $ U $ at 0 to have a charge gap of 3. For $ \lambda_\beta $=1 and $ \omega_\alpha $=$\omega_\beta$=$ \omega_\gamma $=1, the ratio of the increment in $ E_{\rm kin}(\{u\}) $ to that in $ E_\mathrm{tot} $, $ \Delta E_{\rm kin}(\{u\}) / \Delta E_\mathrm{tot} $, increases to about 8$ \times $10$^{-2}$ and $ \Delta E_{\rm kin}(\{v\}) / \Delta E_\mathrm{tot} $ to about 8$ \times $10$^{-3}$ for the distortion amplitude $ u_\mathrm{GS} $ to 0.5. They are one order of magnitude larger than $ \Delta E_{\rm kin}(\{u\}) / \Delta E_\mathrm{tot} < 3 \times 10^{-3} $ and $ \Delta E_{\rm kin}(\{v\}) / \Delta E_\mathrm{tot} < 8 \times 10^{-4} $ in a Mott insulator with dimerization (i.e., a spin-Peierls insulator) even with a larger $ \lambda_\beta $ and the same phonon frequencies. 

We have calculated same quantities for another type of band insulators where the site energies are alternated instead of the transfer integrals. In such a case also, the energy transfer to phonons is generally larger than that in a Mott insulator. This can be understood as follows. In principle, non-interacting electrons in the translation invariant system cannot have a charge gap at half filling. In order for them to have a gap, they must be accompanied by alternation of transfer integrals, site energies, or other potentials, either intrinsically by electron-phonon couplings or extrinsically by $ \delta t $, for instance. Then, the originally inactive (i.e., in the translation invariant system) electronic excitation with wave number $ \pi $ gets optical activity. It means that a certain amount of energy is always transferred to phonons that are coupled with this excitation. Meanwhile, a Mott insulator has a gap even in the translation invariant system, where this excitation remains optically inactive. This suggests that another type of electron-phonon coupling is effective for a Mott insulator to allow large energy transfer to phonons, compared to a band insulator with a similar magnitude of gap. 

\begin{figure}
\includegraphics[height=12cm]{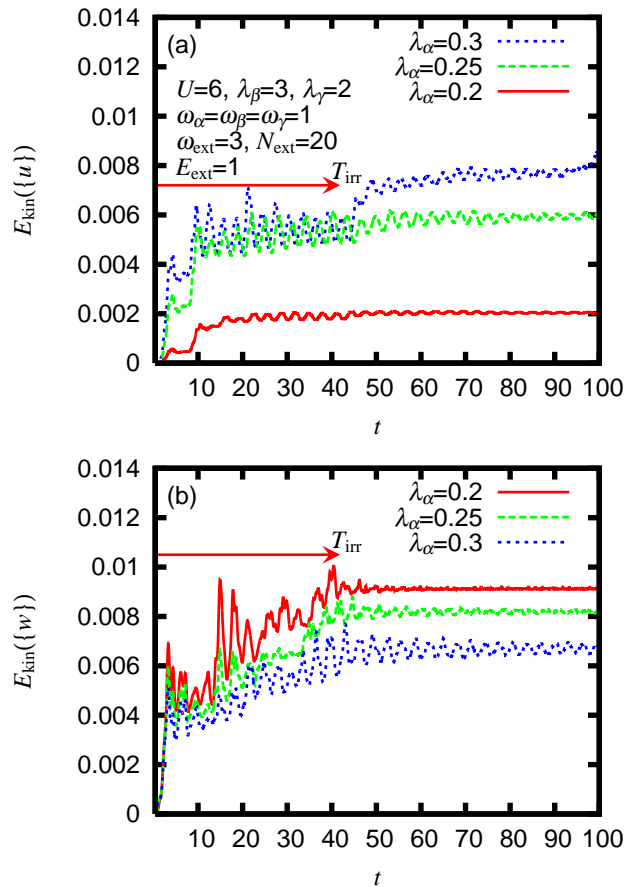}
\caption{(Color online) Time evolutions of kinetic energies of (a) phonons that modulate transfer integrals, $ E_{\rm kin}(\{u\}) $, and (b) phonons that modulate on-site repulsion strengths, $ E_{\rm kin}(\{w\}) $, with different coupling strengths $ \lambda_\alpha $. The other parameters are $ t_0 $=1, $ U $=6, $ \lambda_\beta $=3, $ \lambda_\gamma $=2, $ \omega_\alpha $=$ \omega_\beta $=$ \omega_\gamma $=1, $ \omega_\mathrm{ext} $=3, $ N_\mathrm{ext} $=20, and $ E_\mathrm{ext} $=1. The arrows indicate the pulse duration, $ T_\mathrm{irr} $.
\label{fig:evolution}}
\end{figure}

From now on, we focus on Mott insulators. Although we use a particular set of parameter values to show numerical results, qualitative aspects are general and obtained for other parameter values as well. 
As an example, Figs.~\ref{fig:evolution}(a) and \ref{fig:evolution}(b) show the time evolution of $ E_{\rm kin}(\{u\}) $ and  that of $ E_{\rm kin}(\{w\}) $, respectively. 
The bare phonon frequencies adopted here range from 0.01 to 1, so that their periods range from 6.3 to 628. Thus, the rapid oscillations (of period about 2) are due to either forced or almost resonantly-induced electronic excitations, which are coupled with these phonons. Some structures on larger time scales [of periods about 6 and 12 in Figs.~\ref{fig:evolution}(a) and \ref{fig:evolution}(b)] are also visible. Their appearance is allowed by the couplings with phonons. In the figures, the arrows indicate the pulse duration, $ T_\mathrm{irr} $. After the oscillating electric field is switched off, $ t > T_\mathrm{irr} $, the forced oscillation disappears, but rapid oscillations due to electronic excitations survive. The energy transfer to phonons continues on a large time scale [of the largest period 628 (not shown)]. It becomes conspicuous especially for low (maximum) phonon frequencies. The dependence on the phonon frequencies is discussed later. 

For $ \lambda_\alpha $ values used in these figures, the ground state is in the spin-Peierls phase that has alternating $ u^{(m)}_{j} \propto (-1)^j $. Without this dimerization, the ground state is translation invariant and $ E_{\rm kin}(\{u\}) $=$ E_{\rm kin}(\{v\}) $=0, which is numerically confirmed in the present finite-size system with small $ \lambda_\alpha $. It should be noted here that this is not really the case if phonons are treated quantum-mechanically as in Ref.~\onlinecite{matsueda_prb08}, where the holon and the doublon are dressed with quantum phonons. As long as phonons are treated classically, the wave function is a direct product composed of the electronic and lattice parts where electrons feel the averaged phonon density and phonons feel the averaged electron density so that electron-phonon correlation is missing. Then, the photocarriers cannot be dressed. Nevertheless, we expect that the different characters of the electron-phonon couplings considered here basically survive quantum fluctuations. Unless $ \lambda_\gamma $=0, $ E_{\rm kin}(\{w\}) $ always becomes finite because the optical excitation involves the modulation of double occupancy $ \langle n_{j,\uparrow} n_{j,\downarrow} \rangle $, irrespective of whether the ground state is dimerized or regular. Figure~\ref{fig:evolution}(a) shows that $ E_{\rm kin}(\{u\}) $ increases with $ \lambda_\alpha $. It is because $ \lambda_\alpha $ increases the distortion amplitude $ u_\mathrm{GS} $, which enhances the optical activity of the electronic excitation with wave number $ \pi $ by deviating further from the translation invariance. Figure~\ref{fig:evolution}(b) shows that $ E_{\rm kin}(\{w\}) $ decreases as $ \lambda_\alpha $ increases. It is because $ \lambda_\alpha $ and thus $ u_\mathrm{GS} $ increase the antiferromagnetic spin correlation on the strong bonds (with the larger transfer integrals), which makes the modulation of double occupancy harder. These facts will be made clearer when they are plotted as a function of the distortion amplitude later. 

Not all the energy transfer to phonons is regarded as contributing to dissipation. Some of the transferred energy will flow back into the electronic system and forth into the lattice system, causing oscillations of kinetic energies of phonons. To estimate the component that contributes to dissipation, in other words, to remove the effect of such energy oscillations, we draw a curve, for instance for $ E_{\rm kin}(\{u\}) $, by connecting local minima in time (when it oscillates), measure its value [which is not greater than the bare $ E_{\rm kin}(\{u\}) $] at the observation time, and regard it as the increment in $ E_{\rm kin}(\{u\}) $, $ \Delta E_{\rm kin}(\{u\}) $. 
\begin{figure}
\includegraphics[height=12cm]{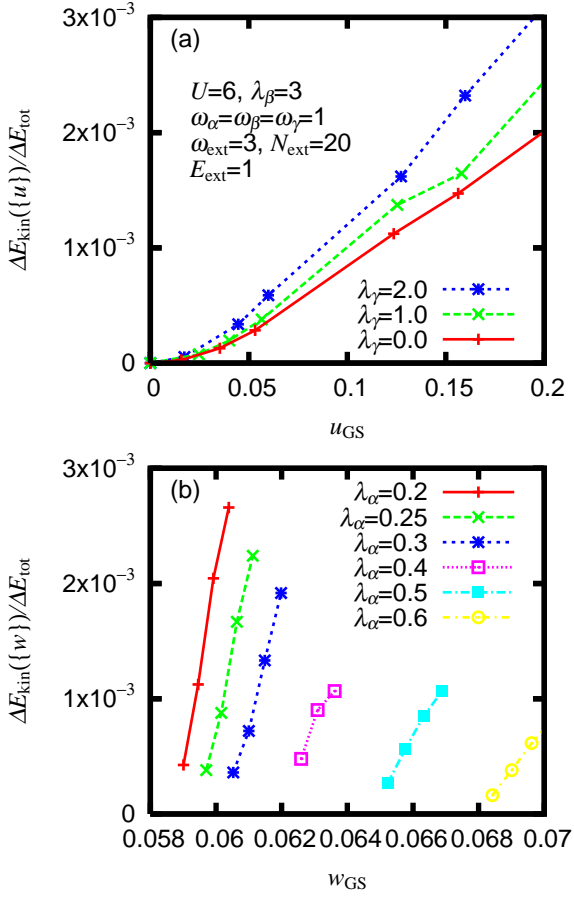}
\caption{(Color online) (a) Ratio of increment in $ E_{\rm kin}(\{u\}) $ to that in $ E_\mathrm{tot} $, as a function of the distortion amplitude $ u_\mathrm{GS} $, with different coupling strengths $ \lambda_\gamma $. (b) Ratio of increment in $ E_{\rm kin}(\{w\}) $ to that in $ E_\mathrm{tot} $, as a function of the distortion amplitude $ w_\mathrm{GS} $, with different coupling strengths $ \lambda_\alpha $. Both are measured at $ t $=100. The parameters are  $ t_0 $=1, $ U $=6, $ \lambda_\beta $=3, $ \omega_\alpha $=$ \omega_\beta $=$ \omega_\gamma $=1, $ \omega_\mathrm{ext} $=3, $ N_\mathrm{ext} $=20, and $ E_\mathrm{ext} $=1.
\label{fig:dist_dependence}}
\end{figure}
Figure~\ref{fig:dist_dependence}(a) shows the ratio of $ \Delta E_{\rm kin}(\{u\}) $ to the increment in the total energy $ \Delta E_\mathrm{tot} $, as a function of the distortion amplitude $ u_\mathrm{GS} $. 
In the classical picture of phonons, unless the ground state has a finite lattice distortion, the lattice remains undistorted even after the photoexcitation, which does not break the translational symmetry. That is, for $ u_\mathrm{GS} $=0, no energy is transferred to $ u $-phonons. As a consequence, $ u_\mathrm{GS} $ is a controlling parameter for the energy transfer to  $ u $-phonons. 
As noted previously, it is a monotonically increasing function. This quantity increases with the coupling $ \lambda_\gamma $ also. The coupling $ \lambda_\gamma $ reduces the on-site repulsion strengths and enhances the superexchange interactions between the spins on the strong bonds, which strengthens the effective coupling between electrons and $ u $-phonons. Figure~\ref{fig:dist_dependence}(b) shows the ratio of $ \Delta E_{\rm kin}(\{w\}) $ to $ \Delta E_\mathrm{tot} $, as a function of the distortion amplitude $ w_\mathrm{GS} $. It is a steeply increasing function above a threshold. The threshold is present because the distortion amplitude $ w_\mathrm{GS} $ defined in Eq.~(\ref{eq:dist_amp_w}) is finite in the $ \lambda_\gamma  \rightarrow 0 $ limit. This quantity is suppressed by the coupling $ \lambda_\alpha $, as explained previously. Namely, the energy flow into $ w $-phonons is suppressed if the dimerization of transfer integrals is large. 

We vary the distributions of phonon frequencies to see their effect on the energy transfer to phonons. 
\begin{figure}
\includegraphics[height=12cm]{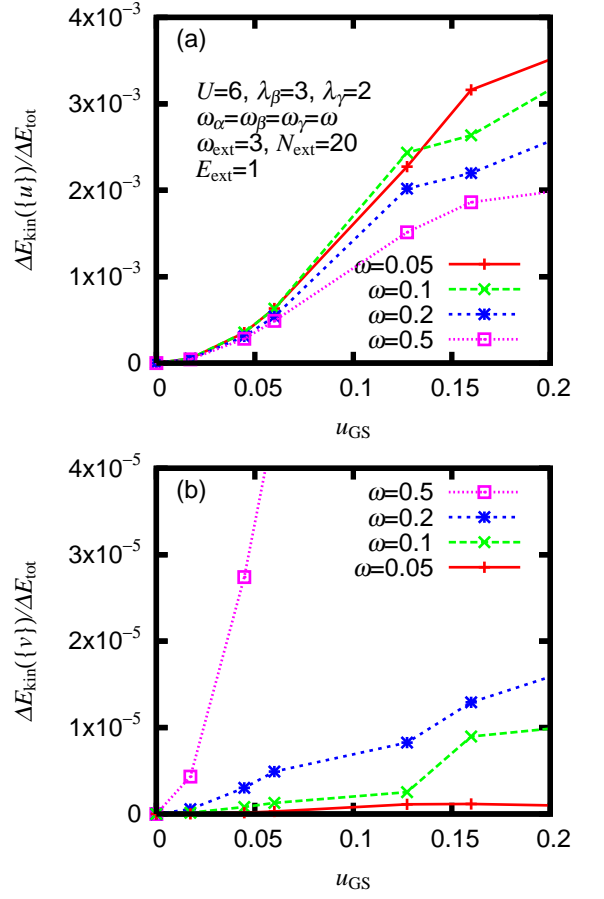}
\caption{(Color online) Ratios of (a) increment in $ E_{\rm kin}(\{u\}) $ to that in $ E_\mathrm{tot} $, and (b) increment in $ E_{\rm kin}(\{v\}) $ to that in $ E_\mathrm{tot} $, as a function of the distortion amplitude $ u_\mathrm{GS} $, with different maximum phonon frequencies $ \omega $=$ \omega_\alpha $=$ \omega_\beta $=$ \omega_\gamma $. Both are measured at $ t $=100. The parameters are $ t_0 $=1, $ U $=6, $ \lambda_\beta $=3, $ \lambda_\gamma $=2, $ \omega_\mathrm{ext} $=3, $ N_\mathrm{ext} $=20, and $ E_\mathrm{ext} $=1.
\label{fig:freq_dependence}}
\end{figure}
Figure~\ref{fig:freq_dependence}(a) shows that the ratio of $ \Delta E_{\rm kin}(\{u\}) $ to $ \Delta E_\mathrm{tot} $ increases with shifting the distributions of phonon frequencies to the low side. Here, the energy ratios are observed at $ t $=100 shortly after the oscillating electric field is switched off. If we decrease the phonon frequencies further, even the lattice vibration of the highest frequency $ \omega_\alpha $ has the period longer than the observation time, leading to decreasing of the energy ratio. However, if we observe it much later so as to allow a few cycles of the fastest lattice vibration before the observation time, the energy ratio tends to increase further. Although we cannot reach the adiabatic limit of $ \omega \rightarrow 0 $ due to numerical limitations, the energy flow into $ u $-phonons seems to become the largest in this limit. This behavior is in contrast to that of $ v $-phonons. Figure~\ref{fig:freq_dependence}(b) shows that the ratio of $ \Delta E_{\rm kin}(\{v\}) $ to $ \Delta E_\mathrm{tot} $ is almost completely suppressed if phonon frequencies are low. Note the difference in the scales of the ordinate axes. This energy ratio increases with shifting the distributions of phonon frequencies to the high side as long as phonon energies are lower than electronic excitation energies. If some phonon energies become comparable with electronic excitation energies, this quantity would show a complex behavior as a function of phonon frequencies. 

These contrastive behaviors of $ u $- and $ v $-phonons are understood in the following way. The dimerization $ u^{(m)}_{j} \propto (-1)^j $ breaks the translation invariance, giving the optical activity to an electronic excitation with wave number $ \pi $. This electronic mode is photoexcited almost resonantly here and transfers energy directly to $ u $-phonons. However, the energy transfer to $ v $-phonons is indirect. This electronic excitation makes the charge densities at even and odd sites different, which applies a force to $ v $-phonons. It then induces oscillations of charge densities and those of $ v $-phonons. The energy flow into thus indirectly coupled $ v $-phonons is facilitated by making different excitations closer in energy. Consequently, the energy flow into $ v $-phonons seems to become largest when their energies are comparable with those of electronic excitations. 

\section{Summary}

The energy flow from a correlated electron system into the environments after photoexcitation is theoretically studied in a one-dimensional half-filled Peierls-Holstein-Hubbard model augmented with another type of electron-lattice couplings that modulate Coulomb repulsion strengths. We consider different types and frequencies of lattice vibrations, which are treated classically. The time-dependent Schr\"odinger equation is numerically solved for the exact many-electron wave functions. The time evolutions of the kinetic energies of different types of phonons are observed and compared with the increment in the total energy. When Mott and band insulators with a similar magnitude of gap are compared in the present half-filled model, the energy transfer to phonons is generally larger in a band insulator than in a Mott insulator. Because the translational symmetry is broken to have a finite gap in the band insulator at half filling, the optical activity is given to an electronic excitation with wave number $ \pi $ and phonons coupled with it. 

In the spin-Peierls phase, the dimerization that alternates transfer integrals gives the optical activity to the electronic excitation with wave number $ \pi $, which allows the energy transfers to phonons that modulate transfer integrals and to phonons that modulate site energies. These energy transfers increase with dimerization amplitude, but their dependences on phonon frequencies are quite different. The transfer to phonons that modulate transfer integrals increases with decreasing phonon frequencies, while that to phonons that modulate site energies increases with increasing phonon frequencies as long as these phonon energies are lower than the electronic excitation energies. This difference originates from the fact that the coupling with the optically active electronic excitation is direct for the former phonons and indirect for the latter phonons. Nevertheless, these energy transfers are generally smaller than those in band insulators. 

The above facts suggest that another type of electron-phonon couplings that modulate Coulomb repulsion strengths are effective in transferring energy to phonons. It should be noted that Coulomb repulsion strengths are generally much larger than the charge gap. Therefore, this type of electron-phonon couplings can be quite strong in principle. They are not usually regarded as substantial because they do not change the symmetry of the electronic state. Their effects can be absorbed into redefinition of Coulomb repulsion strengths as long as the equilibrium properties are concerned. Their effects can be significant in such nonequilibrium conditions as in the photoinduced dynamics. The energy transfer to these phonons always takes place irrespective of whether the ground state is dimerized or regular. They are relatively important when the dimerization is weak. 

In view of the energy transfer to phonons, those which modulate transfer integrals compete with those which modulate on-site repulsion strengths. With increasing dimerization amplitude, the antiferromagnetic spin correlation is strengthened on the strong bonds, making the modulation of double occupancy harder. It is interesting that the electron-phonon couplings that modulate on-site repulsion strengths enhance the energy transfer to phonons that modulate transfer integrals by reducing the on-site repulsion and thus by increasing the superexchange interaction. Above all, electron-phonon couplings that modulate Coulomb repulsion strengths can generally be important when the energy dissipation from the photoexcited state is considered. Although we do not study here, the nearest-neighbor repulsion would also be modulated by relevant phonons and can be quantitatively important. For instance, in transition metal oxides, the repulsion between an electron in a $ d $-orbital and another in an O $ p $-orbital would be strongly modulated by the displacement of the oxygen ion. Because the oxygen ion is light and it can oscillate with high frequency, it would allow quite a rapid decay of the photoexcited state. 

\begin{acknowledgments}
This work was supported by Grants-in-Aid for Scientific Research (C) (Grant No. 19540381), for Scientific Research (B) (Grant No. 20340101), and ``Grand Challenges in Next-Generation Integrated Nanoscience" from the Ministry of Education, Culture, Sports, Science and Technology of Japan.
\end{acknowledgments}

\bibliography{relax08a_2}

\end{document}